\documentstyle[aps,preprint,tighten,epsfig]{revtex}
\def\beq{\begin{equation}}
\def\eeq{\end{equation}}
\def\beqa{\begin{eqnarray}}
\def\eeqa{\end{eqnarray}}
\def\MeV{\nobreak\,\mbox{MeV}}
\def\GeV{\nobreak\,\mbox{GeV}}

\begin{document}
\title{\sc  $g_{NK\Lambda}$ and $g_{NK\Sigma}$  from QCD Sum Rules in the 
$\gamma_5 \sigma_{\mu \nu}$ structure}
\author {M.E. Bracco$^1$\thanks{e-mail: bracco@uerj.br}, 
F.S. Navarra$^2$\thanks{e-mail: navarra@if.usp.br} and
\ M. Nielsen$^2$\thanks{e-mail: mnielsen@if.usp.br}\\
\vspace{0.3cm}
{\it $^1$Instituto de F\'{\i}sica,  Universidade do Estado do Rio de Janeiro}\\
{\it Rua S\~ao Francisco Xavier 524 - 20559-900, Rio de Janeiro, RJ, Brazil}\\
\vspace{0.3cm}
{\it $^2$Instituto de F\'{\i}sica, Universidade de S\~{a}o Paulo}\\
{\it C.P. 66318,  05389-970 S\~{a}o Paulo, SP, Brazil}}
\maketitle
\vspace{1cm}

\begin{abstract}
The $NKY$ coupling constant for $ Y = \Lambda $ and $ \Sigma $ 
 is evaluated in a QCD sum rule calculation. 
We study the Borel sum rule for the three
point function of one pseudoscalar one nucleon and one hyperon current
up to order six in the operator product expansion. The Borel transform
is performed with respect to the nucleon and hyperon momenta, which 
are taken to be equal. We discuss and extend the result of a previous 
analysis in the $\rlap{/}{q}i\gamma_5$  structure and compare it with the 
result obtained with the use of the $\gamma_5 \sigma_{\mu \nu}$ structure. 
We find that the effect of pole-continuum transitions is very important 
in the $\rlap{/}{q}i\gamma_5$  structure and that it changes completely
results obtained in previous analysis.
\\
PACS numbers 13.75.Gx,~~12.38.Lg,~~11.55.Hx
\\

\end{abstract}

\vspace{1cm}

Since already long time ago QCD sum rules (QCDSR) \cite{svz} have been used 
to determine  
hadronic coupling constants. The most studied case is the pion-nucleon
 coupling 
\cite{rry,sh,bk,klo1,klo2}. In Ref.\cite{rry} this coupling 
constant was calculated using two different approaches, one based on the
three-point function (three interpolating fields sandwiched between vacuum
states), and one based on the two-point function (two interpolating nucleon 
fields sandwiched between the vacuum and the pion states). In both calculations
the continuum contributions were neglected. In Ref.\cite{sh} the two-point
function calculation of Ref.\cite{rry} was improved by including higher
order terms in the operator product expansion (OPE) as well as a perturbative
estimate of continuum contributions, but  the soft-pion limit was  kept. 
More recently, Birse and Krippa \cite{bk} considered
the two-point correlator beyond the soft-pion limit and they included also
the single pole contribution associated with the transition $N\rightarrow 
N^*$ \cite{is}. However, in Refs.\cite{klo1,klo2}, it was pointed out that
there is a coupling scheme dependence (dependence on how one models the
phenomelogical side either using a pseudoscalar (PS) or pseudovector (PV) 
coupling scheme) in the previous calculations. In order to avoid this scheme
dependence the $\gamma_5\sigma_{\mu\nu}$ Dirac structure was considered.   
In this structure, the phenomenological side has a common double pole 
structure,  independent of the PS or PV coupling schemes. 

There are also works addressing strange couplings, 
like  $ g_{NK\Lambda} $, $g_{NK\Sigma}$, $g_{\pi\Lambda\Sigma}$ and 
$g_{K\Sigma\Xi}$ \cite{cch,sc} and the charm coupling $g_{ND\Lambda_c}$
\cite{nn}, all of them based on the three-point function.
In this work we shall revisit the calculation of the coupling constant 
$g_{NKY}$ in the framework of QCD sum rules (QCDSR) considering the Dirac 
structure suggested in \cite{klo1,klo2}.

In order to calculate the $NKY$ coupling constant using
the QCDSR we consider the three-point function
\begin{equation}
A(p,p^\prime,q)=\int d^4x \, d^4y \, \langle 0|T\{\eta _{Y}(x)
j_5(y)\overline{\eta }_N(0)\}|0\rangle  
\, e^{ip^\prime x} \, e^{-iqy}\; , 
\label{cor}
\end{equation}
constructed with two baryon currents, $\eta _{Y}$ and $\eta _N$,
for hyperon and the nucleon respectively, and the pseudoscalar meson
$K$ current, $j_5$, given by \cite{rry,cch,ioffe,rry2,cpw}
\beq
{\eta }_{\Lambda}= \sqrt{\frac{2}{3}}
\varepsilon _{abc}[({u}_a^TC\gamma _{\mu}{s}_b)\gamma_5 \gamma^{\mu}d_c 
- ({d}_a^TC\gamma _{\mu}{s}_b)\gamma_5 \gamma^{\mu}u_c]
\label{lambcur}
\eeq
\beq
{\eta }_{\Sigma^0}= \sqrt{2}
\varepsilon _{abc}[({u}_a^TC\gamma _{\mu}{s}_b)\gamma_5 \gamma^{\mu}d_c 
+ ({d}_a^TC\gamma _{\mu}{s}_b)\gamma_5 \gamma^{\mu}u_c]
\label{sigcur}
\eeq
\beq
{\eta }_N= \varepsilon _{abc}({u}_a^TC\gamma^\mu{u}_b)\gamma_5\gamma_\mu d_c 
\; ,
\eeq
\beq
j_5=\overline{s} i\gamma_5u \; ,
\eeq
where $s$, $u$ and $d$  are the strange, up and down quark fields
respectively and  $C$ is the charge conjugation matrix.

Due to restrictions from Lorentz, parity and charge conjugation invariance
the general expression for $A(p,p^\prime,q)$ in Eq.(\ref{cor}) has the form
\beqa
A(p,p^\prime,q)&=&F_1(p^2,p'^2,q^2)i\gamma_5 +F_2(p^2,p'^2,q^2)\rlap{/}{q}
i\gamma_5
\nonumber\\*[7.2pt]
&+&F_3(p^2,p'^2,q^2)\rlap{/}{P}i\gamma_5+ F_4(p^2,p'^2,q^2)\sigma^{\mu\nu}
\gamma_5 q_\mu p^\prime_\nu \; ,
\label{struc}
\eeqa
where $q=p^\prime-p$ and $P=(p+p^\prime)/2$.

The calculations of the coupling constants based on the three-point function
have concentrated on the $\rlap{/} {q}i\gamma_5$ structure 
\cite{rry,cch,sc,nn} developing sum rules for $F_2$. 
In the case of $g_{\pi N N}$,  Shiomi and Hatsuda 
\cite{sh} considered the $ i\gamma_5 $ structure whereas Birse and Krippa 
\cite{bk} used the $ \rlap{/}{P}i\gamma_5 $ structure. 
In principle any of these structures can 
be used to calculate the coupling constant and the sum rules should yield the 
same result. However, each sum rule could have uncertainties due to the 
truncation in the OPE side and different contributions from the continuum.
Therefore, depending on the Dirac structure we can obtain different results 
due to the uncertainties mentioned above. The tradicional way 
to control these uncertainties, and therefore to check the reliability of 
the sum rule, is to choose the Borel window appropriately (in such a way
that the continuum contribution is less than $40\sim50\%$), and evaluate the
stability of the result as a function of the Borel mass.

As mentioned before, in Refs.\cite{klo1,klo2} it was pointed out that a better 
determination of    $g_{\pi N N}$ can be done with the 
help of the $\gamma_5\sigma_{\mu\nu}$ structure, since  
this structure is independent of the effective models employed in the 
phenomenological side and it gets a smaller contribution from the
single pole term coming from $N\rightarrow N^*$ transition.
Motivated by this finding we shall calculate $g_{NKY}$  in this structure 
and compare our results with previous QCDSR 
estimates \cite{cch,sc}.

In the phenomenological side the different Dirac structures appearing in 
Eq.(\ref{struc}) are obtained by the consideration of the $Y$ and 
$N$ states contribution to the matrix element  in Eq.~(\ref{cor}):
\begin{equation}
\langle 0|\eta _{Y}|Y(p^\prime)\rangle\langle Y
(p^\prime)|j_5|
N(p)\rangle\langle N(p)|\overline\eta_N|0\rangle \; ,
\label{j5}
\end{equation}
where the matrix element of the pseudoscalar current defines the pseudoscalar 
form-factor
\begin{equation}
\langle Y(p^\prime)|j_5|N(p)\rangle=g_P(q^2)\overline{u}(p^\prime)i
\gamma_5
u(p)\; ,
\label{ps}
\end{equation}
where $u(p)$ is a Dirac spinor and $g_P(q^2)$ is related to $g_{NDY}$
through the relation \cite{rry}
\beq
g_P(q^2)={m_K^2 f_K\over m_q}{g_{NKY}\over q^2-m_K^2} \; ,
\label{g}
\eeq
where $m_K$ and $f_K$ are the kaon mass and decay constant and
$m_q$ is the average of the quark masses: $(m_u+m_s)/2$. 
 
The other matrix elements contained in Eq.(\ref{j5}) are
of the form
\begin{eqnarray}
\langle 0|\eta _{Y}|Y(p^\prime)\rangle &=&\lambda_{Y} 
u(p^\prime) 
\label{lamc}
\\
\langle N(p)|\overline\eta_N|0\rangle &=& \lambda_N
\overline{u}(p)\; ,
\label{lam}
\end{eqnarray}
where $\lambda_{Y}$ and $\lambda_N$ are
the couplings of the currents with the respective hadronic states.

Saturating the correlation function Eq.(\ref{cor}) with
$Y$ and $N$ intermediate states, and using Eqs.~(\ref{j5}),
(\ref{ps}), (\ref{g}),  (\ref{lamc}) and (\ref{lam}) we get 
\begin{eqnarray}
A^{(phen)}(p,p^\prime,q) &=&\lambda_{Y}\lambda_N {m_K^2 f_K
\over m_q}{g_{NKY}\over q^2-m_K^2}
{(\rlap{/}{p^\prime}+M_{Y})\over p'^2-
M_{Y}^2} i\gamma_5{(\rlap{/}{p}+M_N)\over p^2-M_N^2} + 
\mbox{higher resonances}\; ,
\label{aphen}
\end{eqnarray}
which can be rewritten as
\begin{eqnarray}
A^{(phen)}(p,p^\prime,q) &=&\lambda_{Y}\lambda_N {m_K^2 f_K
\over m_q}{g_{NKY}\over q^2-m_K^2}
{1\over p'^2-M_{Y}^2}{1\over p^2-M_N^2}\left[(M_{Y}M_N-p.
p^\prime)i\gamma_5\right.
\nonumber \\*[7.2pt]
&+&\left. {M_{Y}+M_N\over2}\rlap{/}{q}i\gamma_5 - (M_{Y}-M_N)
\rlap{/}{P}i\gamma_5-\sigma^{\mu\nu}\gamma_5q_\mu p^\prime_\nu \right] + 
\mbox{higher resonances}\; ,
\label{ficor}
\end{eqnarray}
where we clearly see all the Dirac structures present in Eq.(\ref{struc}).

We will write a sum rule for the  $\sigma^{\mu\nu} \gamma_5 q_\mu 
p^\prime_\nu$ structure. As we are interested in the value of the
coupling constant at $q^2=0$, we will make a Borel transform to
both $p^2={p^\prime}^2\rightarrow M^2$. In Eq.~(\ref{ficor}) higher
resonances refers to pole-continuum transitions as well as pure continuum 
contribution. The pure continuum contribution 
will be taken into account as usual through the standard form
of Ref.\cite{ioffe}.

As in any QCD sum rule calculation, our goal is to make a match between the 
two representations of the correlation function (\ref{cor}) at a certain 
region of $M^2$: the OPE side and the phenomenological side.

In the OPE side only even dimension operators contribute to the 
$\sigma^{\mu\nu} \gamma_5 q_\mu p^\prime_\nu$ structure, since the dimension 
of Eq.(\ref{cor})
is four and $q_\mu p^\prime_\nu$ take away two dimensions.

Following Refs.~\cite{cch,sc} we will neglect $m_K^2$ and $ m_s^2$ in the 
denominators and, 
consequently, only terms proportional to $ 1/ q^2$ will contribute to the sum 
rule.

For the $\Lambda$ hyperon the diagrams that contribute up to dimension six
are shown in Fig. 1. The 
lowest dimension operator is the mass times the quark condensate with 
dimension four (Fig. 1a). Since we are neglecting the 
light quark masses, only terms proportional to $m_s \langle\overline{s}s
\rangle $ will appear. This term gives, after a Borel tranformation  with 
respect to $P^2=-p^2=-{p^\prime}^2$: 
\beq
\left[\tilde{F_4}(M^2,q^2)\right]_{a} = - \sqrt{\frac{2}{3}} 
\frac{m_s \langle\overline{s}s\rangle}{16\pi^2} \frac{M^2}{q^2} E_0^{\Lambda}
\label{d4a}
\eeq
with $E_0^{\Lambda}=1-e^{-s_{\Lambda}/M^2}$.
In the above equation $\tilde{F_4}$ stands for the Borel transformation
of the amplitude $F_4$ and $s_{\Lambda}$ gives the continuum threshold for 
$\Lambda$.

The next contribution comes from the diagrams involving dimension 6 operators
of the type  $\langle\overline{q}q\overline{q}q\rangle$ ($\simeq \langle
\overline{q}q\rangle ^2$) shown in Figs.~1b and 1c.  
The expressions for these contributions are
\beq
\left[\tilde{F_4}(M^2,q^2)\right]_{b} = - \frac{4}{3} \sqrt{\frac{2}{3}} 
\frac{\langle\overline{q}q\rangle^2}{q^2} \, ;
\label{d6b}
\eeq
and 
\beq
\left[\tilde{F_4}(M^2,q^2)\right]_{c} = - \frac{4}{3} \sqrt{\frac{2}{3}} 
\frac{\langle\overline{q}q\rangle \, \langle\overline{s}s\rangle}{q^2} \, .
\label{d6c}
\eeq

There would be still another class of contributions to the OPE side 
related to the four quark processes discussed in Ref. \cite{jk}. In Ref. 
\cite{jk} it was shown that four quark processes with nonlocal condensates
in the three-point function approach are equivalent to the susceptibilities
processes in the two-point formalism \cite{hen} and, therefore, should
be included when calculating low-momentum transfer processes in the QCD 
sum rule
approach. However, this term does not contribute to the $\sigma^{\mu\nu} 
\gamma_5 q_\mu p^\prime_\nu$ structure.

The Borel transformation of the phenomenological side gives
\beq
\left[\tilde{F_4}(M^2,q^2)\right]_{phen}=-\lambda_{\Lambda}\lambda_N 
{ m_K^2 f_K\over m_q}{g_{NK\Lambda}\over q^2}{
1\over M_{\Lambda}^2-M_N^2}(e^{-M_N^2/M^2}-e^{-M_{\Lambda}^2/M^2})\;+\; 
\cdot\cdot\cdot\;\; ,\label{fen}
\eeq
where the dots include now only the contribution from the 
unknown single pole terms since the pure continuum contribution has 
already been incorporated in the OPE side, through the factor $E_0$ in 
Eq.~(\ref{d4a}).

For $\lambda_{\Lambda}$ and $\lambda_N$ we use the values obtained from
the respective mass sum rules for the nucleon  and for $\Lambda$ 
\cite{rry,cch,ioffe}:

\beq
|\lambda_N|^2e^{-M_N^2/M^2} 2 (2\pi)^4
= M^6 E_2^N + {4\over3} a^2\; ,
\label{rsn}
\eeq
\beqa
|\lambda_{\Lambda}|^2e^{-M_{\Lambda}^2/M^2} 2 (2\pi)^4 = M^6 E_2^{\Lambda} + 
\frac{2}{3} a m_s (1-3 \gamma) M^2 E_0^{\Lambda} +b M^2 E_0^{\Lambda} 
 +\frac{4}{9} a^2 (3+4 \gamma)\; ,
\label{rslc}
\eeqa
where $a=-(2\pi)^2 \langle\overline{q}q\rangle \simeq 0.5 \, \GeV^3$,  
$b=\pi^2 \langle \alpha_sG^2/\pi\rangle \simeq 0.12\,\GeV^4$
and $\gamma = \langle\overline{q}q\rangle/ \langle\overline{s}s\rangle -1 
\simeq -0.2$. In the above expressions 
$E_2^{\Lambda(N)}=1-e^{-s_{\Lambda(N)}/M^2}(1+s_{\Lambda(N)}/M^2
+s_{\Lambda(N)}^2/(2M^4))$, 
with $s_N$ being the continuum threshold for the nucleon. The $E_i$ factors
in Eqs.~(\ref{rsn}) and (\ref{rslc}) accounts for the continuum contribution.

In order to obtain $g_{NK\Lambda}$ we identify Eq.~(\ref{fen}) with the 
sum of Eqs.~(\ref{d4a}), (\ref{d6b}), (\ref{d6c}). We obtain \cite{bk,klo2,is}:
\beqa
g_{NK\Lambda}+AM^2&=&
-\sqrt{2\over3}{1\over\tilde{\lambda}_{\Lambda}\tilde{\lambda}_N}{m_q\over 
m_K^2 f_K}{M_{\Lambda}^2-M_N^2\over e^{-M_N^2/M^2}-e^{-M_{\Lambda}^2/M^2}}\
a\left[{m_s\gamma\over4}M^2E_0^{\Lambda}\right.
\nonumber \\*[7.2pt]
&-&\left.{4\over3}a(1+\gamma)\right] \; ,\label{sr}
\eeqa
where $\tilde{\lambda}_{N(\Lambda)}=(2\pi)^2\lambda_{N(\Lambda)}$ and $A$
denotes the contribution from the unknown single pole term coming from
$N\rightarrow N^*$ transition which is not suppressed by the Borel 
transformation \cite{bk,klo2,is}. Using Eqs.~(\ref{rsn}), (\ref{rslc})
we solve  Eq.~(\ref{sr}) for  $g_{NK\Lambda}$ by fitting its
right hand side (RHS) by a straight line, in the appropriate Borel window. 
Since Eqs.~(\ref{rsn}) and (\ref{rsls}) determine only the absolute 
value of $\lambda_N$ and $\lambda_{\Lambda}$ we can not determine the 
sign of $g_{NK\Lambda}$.

In this calculation there are
some ingredients which are heavily constrained by  theoretical or 
phenomenological analyses. The quark condensate is taken to be  
$\langle\overline{q}q\rangle\,=\,-(0.23)^3\,\GeV^3$.   
The continuum  thresholds  appearing in the $E_i$ factors are chosen to be  
$s_{\Lambda}\,=\,(M_{\Lambda}+0.5)^2\,\GeV^2$ and 
$s_N\,=\,(M_N+0.5)^2\,\GeV^2$. The hadron 
masses are $M_N\,=\,0.938\,\GeV$, $M_{\Lambda}\,=\,1.150\,\GeV$ and 
$m_K\,=\,0.495\,\GeV$. The strange quark mass is taken to be $m_s\,=\,150$ 
\, MeV and the kaon decay constant is $f_K={160\over\sqrt{2}}\MeV
\simeq113\MeV$.                    
The relevant Borel mass here is $M\simeq\frac{M_N+M_{\Lambda}}{2}$ and 
we analyse the sum rule in the interval $0.8 \leq M^2 \leq 1.6$ 
\GeV$^2$ where the continuum contribution is always smaller than 50\% of
the total OPE. 

In Fig.~2 we show the RHS of Eq.~(\ref{sr}) as a function of the Borel
mass squared (thick solid line). We show the results in a broader Borel
range than discussed above to show that our conclusions are not very 
constrained by the Borel window used. To check the sensitivity of our result 
on the continuum contribution, we have increased the continuun thresholds
as: $s_{\Lambda}\,=\,(M_{\Lambda}+0.7)^2\,\GeV^2$ and 
$s_N\,=\,(M_N+0.7)^2\,\GeV^2$, and plotted the corresponding result as the 
thin line in the same figure. As a first sign it seems that the result is
very sensitive to the continuum thresholds. However, as the value of the
coupling constant is obtained by the extrapolation of the line to $M^2=0$,
we imediately see that both curves lead to approximately the same result. 
Indeed, fitting the QCDSR result to a straight line we get

\beq
{|g_{NK\Lambda}|} \,=\,2.37\,\pm\,0.09 \; ,
\eeq
where the error was only estimated by using the two different thresholds.
The value of $A$ is very different in both curves ($0.35(-0.26)$ for the
thick(thin) line). However, in both curves it is very small showing that
the single pole contribution is not very important in this structure, in
agreement with the results in Ref.\cite{klo2}.

As a straightforward extension of our calculation we compute now the coupling 
constant $g_{NK\Sigma}$ also in the $\gamma_5 \sigma_{\mu \nu}$ structure. 
Phenomenological analyses \cite{bonn} 
indicate that this coupling constant should be much smaller 
than $g_{NK\Lambda}$. Moreover this quantity was also computed in 
Refs.\cite{cch,sc} in the $\rlap{/}{q}i\gamma_5$ structure  and therefore
both results can be compared.

From Eqs.~(\ref{lambcur}) and (\ref{sigcur}) we see that apart from an overall
 numerical factor the only difference between $\eta_{\Lambda}$ and 
$\eta_{\Sigma}$ is the sign change. This sign inversion introduces 
cancelations and the final number of terms contributing 
to $F_4$ in the OPE side is smaller. Moreover,  since we take $q^2 \rightarrow
 0$ only the diagram of Fig.~1a contributes:
\beq
\left[\tilde{F_4}(M^2,q^2)\right]_{a} =  \sqrt{2} 
\frac{m_s \langle\overline{s}s\rangle}{16\pi^2} \frac{M^2}{q^2} E_0^{\Sigma}
\label{d4as}
\eeq
The phenomenological side is similar to Eq.~(\ref{fen}) with the replacements 
$\lambda_{\Lambda} \rightarrow \lambda_{\Sigma}$, $M_{\Lambda} \rightarrow 
M_{\Sigma}$ and
$g_{NK\Lambda} \rightarrow g_{NK\Sigma}$. For the coupling factor 
$\lambda_{\Sigma}$ we take 
\cite{rry,cch}:
\beqa
|\lambda_{\Sigma}|^2e^{-M_{\Sigma}^2/M^2} 2 (2\pi)^4 = M^6 E_2^{\Sigma}  
-2a m_s (1+\gamma) M^2 E_0^{\Sigma} +b M^2 E_0^{\Sigma} +\frac{4}{3} a^2 
\label{rsls}
\eeqa
with the same definitions used before and with $M_{\Sigma}=1.189\GeV$.
The final expression for the sum rule is:
\beq
g_{NK\Sigma}+BM^2=
\sqrt{2}{1\over\tilde{\lambda}_{\Sigma}\tilde{\lambda}_N}{m_q\over 
m_K^2 f_K}{M_{\Sigma}^2-M_N^2\over e^{-M_N^2/M^2}-e^{-M_{\Sigma}^2/M^2}}\
{am_s\gamma\over4}M^2E_0^{\Sigma} \; .\label{srsi}
\eeq

In Fig.~3 we plot the RHS of Eq.~(\ref{srsi}) as a function of the Borel
mass squared for $s_{\Sigma}\,=\,(M_{\Sigma}+0.5)^2\,\GeV^2$ and 
$s_N\,=\,(M_N+0.5)^2\,\GeV^2$ (thick solid line) and for
$s_{\Sigma}\,=\,(M_{\Sigma}+0.7)^2\,\GeV^2$ and $s_N\,=\,(M_N+0.7)^2\,
\GeV^2$ (thin solid line). In this case the sensitivity to the continuum
threshold and the contribution from single pole are even smaller. 
Fitting the QCDSR results to a straight line we get:

\beq
{|g_{NK\Sigma}|} \,=\,0.025\,\pm\,0.015\; .
\eeq

The results obtained in \cite{cch} are :
\beq
{|g_{NK\Lambda}|} \,=\,6.96\,\,\,\,\,\mbox{and}\,\,\,\,\,
{|g_{NK\Sigma}|} \,=\,1.05 \; .
\eeq

However, the results in Ref.\cite{cch} were obtained without considering 
continuum contribution and they are shown as the dot-dashed lines in Figs.~2
and 3. Once the continuum contribution is included, through the usual $E_i$ 
factors, the behaviour of the sum rule as a function of the Borel mass 
changes drastically, as can be seen by the dashed line in Figs.~2 and 3.
In particular, both  
$g_{NK\Lambda}$ and  $g_{NK\Sigma}$ become approximately linear functions 
of $M^2$, showing the importance of the pole-continuum  
contribution in this structure. In Ref.\cite{sc} it was found that,  
including the continuum contribution
in the results of Ref.\cite{cch}, one obtains
${|g_{NK\Lambda}|}=
8.34$ and ${|g_{NK\Sigma}|}=1.26$. This result is, however, misleading, 
since
the sum rule was not analyzed as a function of the Borel mass, and  
the single pole contributions were not included.

Fitting the RHS  of the sum rule results on the structure $\rlap{/}{q}i
\gamma_5$ \cite{cch} (including the 
continuum contribution) to a straight line one gets
\beq
{|g_{NK\Lambda}|}=1.5\pm0.3\,\,\,\,\, \mbox{and}\,\,\,\,\,
{|g_{NK\Sigma}|}=0.25\pm0.05\; ,
\eeq
where the errors are again evaluated only by considering the two different
continuum thresholds.

As in Ref.\cite{klo2} we find out that we can obtain very different results 
for the coupling constants depending on the structure considered. Of
course the procedure used here to extract the coupling constant (fitting the
QCDSR result to a straight line in a given Borel window and extrapolating it
to $M^2=0$) is more reliable when the single pole term is small. Therefore,
the results obtained for the structure $\rlap{/}{q}i\gamma_5$ may contain big 
errors since the single pole contribution to this structure is very strong, as
can be seen by the dashed lines in Figs.~2 and 3. On the other hand, we may
say that the results on the structure $\gamma_5 \sigma_{\mu \nu}$, 
analysed here, are not contaminated by the single pole transitions and its
extraction with the method used here is more reliable.

As a final remark we note that the values for the coupling constants obtained 
here in both structures considered, are not in agreement with the
exact SU(3) symmetry. In this limit there are two independent couplings
of pseudoscalar mesons too the baryon octet, usually denoted by $F$ and $D$,
corresponding to antisymmetric and symmetric combinations of the octet 
fields. The SU(3) symmetry, using de Swart's convention \cite{swa}, predicts
\beqa
g_{NK\Lambda}&=& -{1\over\sqrt{3}}(3-2\alpha_D)g_{\pi NN}\; ,
\nonumber\\
g_{NK\Sigma}&=& (2\alpha_D-1)g_{\pi NN}\; ,
\label{deswa}
\eeqa
where $\alpha_D=D/(D+F)$. Taking $\alpha_D=0.64$ \cite{rat} we get from
Eq.~(\ref{deswa}): $|g_{NK\Lambda}/g_{NK\Sigma}|=3.55$. Therefore, our
results show a huge breaking of SU(3) symmetry.
 
\vspace{1cm}

\underline{Acknowledgements}: This work has been supported by FAPESP, CAPES
and CNPq.  We would like to warmly thank M.K. Banerjee, T.D. Cohen and 
E. Henley for 
instructive discussions. M.N. would like to thank the Institute for Nuclear 
Theory at the University of Washington for its hospitality and financial 
support during her stay in Seattle.

\newpage
\noindent
\begin{center}
{\Large\bf Figure Captions}\\
\end{center}
\vskip 2cm
\begin{itemize}

\item[{\bf Fig. 1}] Diagrams that contribute to the OPE side for $g_{NKY}$.

\item[{\bf Fig. 2}] $g_{NK\Lambda}$ as a function of the squared Borel mass
$M^2$ for the $\gamma_5\sigma_{\mu\nu}$ structure (solid line) and for the
$\rlap{/}{q}\gamma_5$ structure with (dashed line) and without (dot-dashed 
line) continuum contibutions. The thick lines are obtained using the continuum
thresholds given by: $s_{\Lambda}\,=\,(M_{\Lambda}+0.5)^2\,\GeV^2$ and 
$s_N\,=\,(M_N+0.5)^2\,\GeV^2$, while for the thin lines lines we used
$s_{\Lambda}\,=\,(M_{\Lambda}+0.7)^2\,\GeV^2$ and 
$s_N\,=\,(M_N+0.7)^2\,\GeV^2$.

\item[{\bf Fig. 3}] Same as Fig.~2 for $g_{NK\Sigma}$.

\end{itemize}
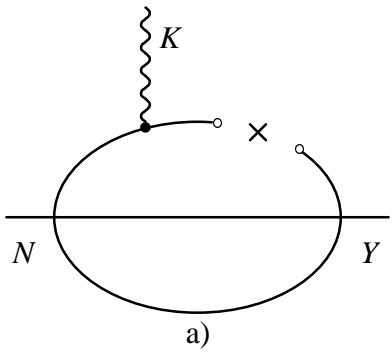
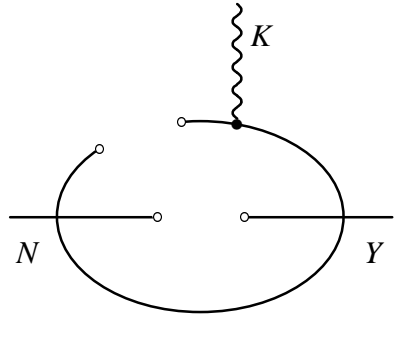
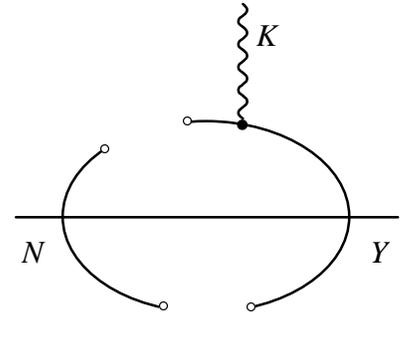
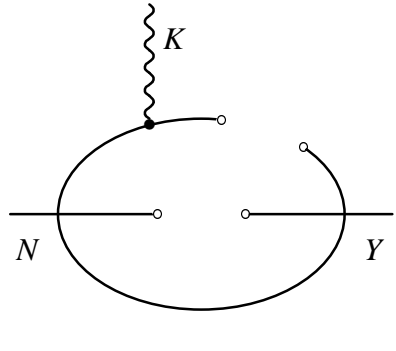
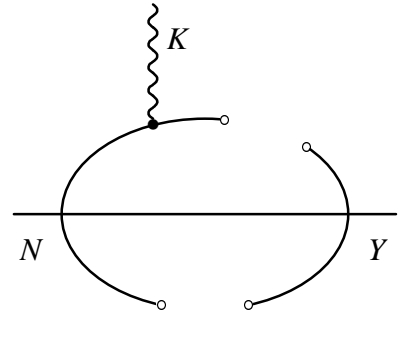
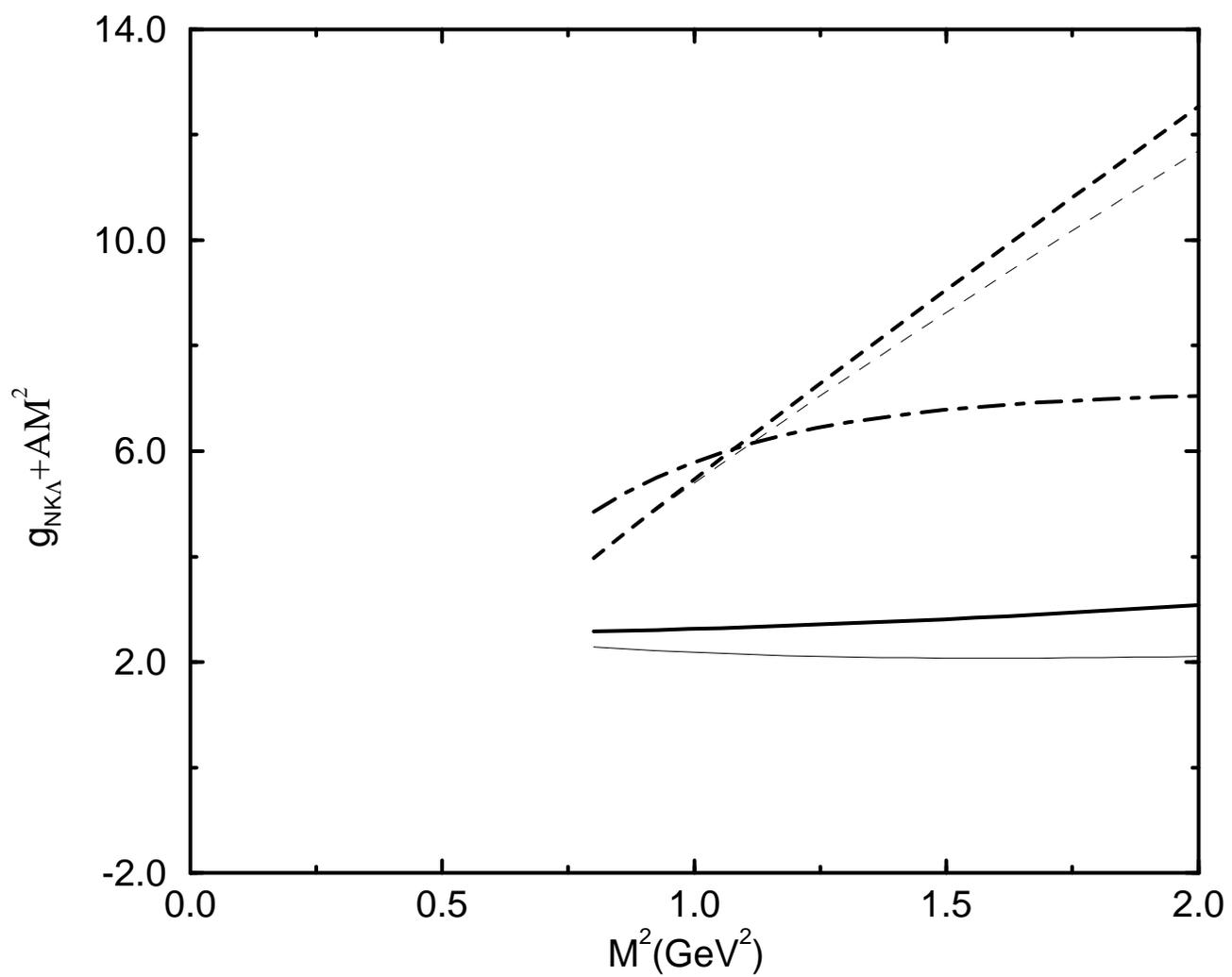
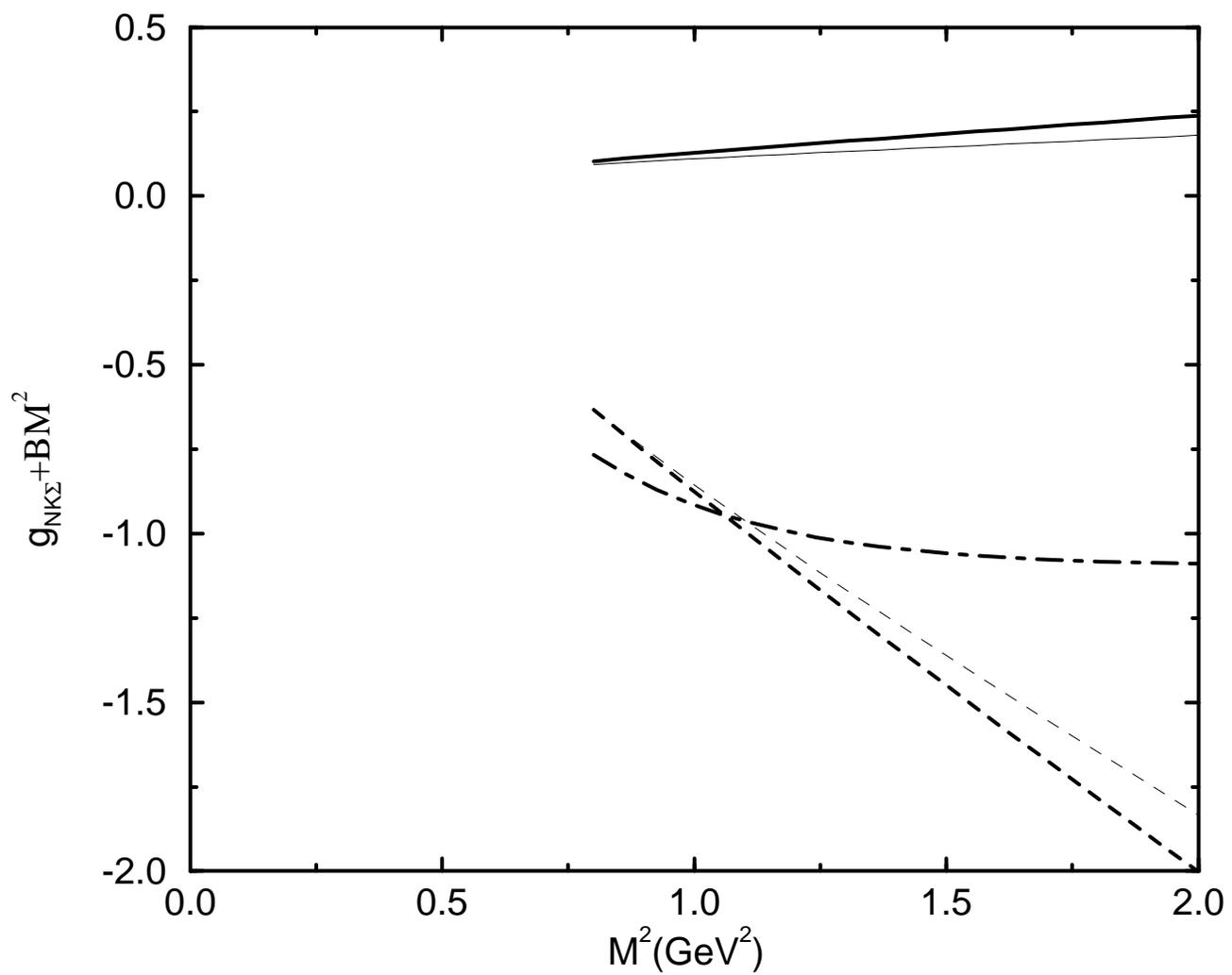
\end{document}